\documentstyle[aps,prl,preprint, tighten]{revtex} 
\begin{document} 
\draft
\preprint{IMSc./2000/08/47}
\title{Quantum gravity as a theory of quantized area bits fitting together}
\author{H.S.Sharatchandra \thanks{e-mail:sharat@imsc.ernet.in}} 
\address{The Institute of Mathematical Sciences, C.I.T Campus, Taramani P.O.,
Chennai 600 113.}
\addtocounter{footnote}{1}
\author{H.Gopalkrishna Gadiyar \thanks{e-mail:gadiyar@au-kbc.org}}
\address{AU-KBC Center for Internet $\&$ Telecom Technologies, M.I.T. Campus 
of Anna University, Chromepet,  Chennai 600 044.}
\maketitle
\begin{abstract} 
Non-Abelian Gauss law is interpreted in terms of area bits described in a local
frame which fit together into closed surfaces and the  Non-Abelian Stokes law 
in terms of length bits described in a local frame which fit together into closed loops. 
A new equation relating the area variables and the phase space variables ( or
equivalently, angular momentum variables of the lattice Yang-Mills theory 
and phase space variables of the continuum theory) is obtained. Canonical 
quantization applied to these variables implies area quantization. A complete 
orthonormal basis of states satisfying the Gauss constraint 
is obtained.It has the interpretation of quantized area bits with undefined
orientations and edges but fitting together into closed surfaces. 
\end{abstract}

Quantum gravity has posed many intriguing puzzles which have not been fully 
unraveled yet. Einstein gravity is a non-renormalizable theory with a dimensionful
coupling constant $\ell_p=\sqrt {\hbar G c^{-3}}$, the Planck length. It is a 
popular conjecture that this length may play a fundamental role in 
Nature curing the divergences of quantum gravity and perhaps of matter fields too. 
Semi-classical analysis of the black holes shows that the area of the horizon 
has a mysterious significance and it has even been speculated that it may be 
quantized \cite{bek}. In the loop formulation of gravity variables \cite{rov} 
for area, volume etc. have been proposed.
It has been argued that they have non-trivial commutation relations on quantization,
leading to quantized values and giving space time a non-commutative geometry.  

In this paper we analyze the connection  formalism \cite{ash} for gravity (in the real
connection formulation \cite{bar}) in a different way. We are able to solve the
Gauss law constraint explicitly in the quantum theory and obtain a complete
orthonormal set of basis states. This clarifies the precise sense in which 
classical geometry is relevant to quantum gravity. It also brings out the 
precise sense in which area is quantized.

Consider Barbero's  formulation \cite{bar} of classical gravity with 
a real $SU(2)$
connection. The 3-dimensional space is taken to be an oriented, 
analytic manifold $\Sigma$ 
described by arbitrary local coordinates $\vec{x}=\{ x^a \}$.The configuration space 
consists of all smooth $SU(2)$ connections $\vec{A}_a(\vec{x})= \{ A_a^i(\vec{x}) \}$ 
satisfying appropriate boundary conditions.Here $i=1,2,3$ is the internal
index (i.e. the label for the adjoint representation of  $SU(2)$) and $a=1,2,3$ is the 
tangent space index.  The momentum  space is the
space of smooth vector densities $\vec{E}_a(\vec{x})= \{ E^{ai}(\vec{x}) \}$ of weight 
one on $\Sigma$. The Poisson
brackets are
\begin{equation}\label{ccr}
[A_a^i(\vec{x}),E^{bj}(\vec{y})]_{PB}=
\delta_{ij} \delta_{ab} \delta(\vec{x} - \vec{y})
\end{equation}
Other brackets are zero.Only part of this phase space satisfying
seven first class constraints is physically accessible and specify the theory.
The first three are the non-Abelian Gauss law 
~
\begin{equation}\label{gauss}
D_{a}[\vec{A}] \vec{E}^{a}(\vec{x})=0.
\end{equation}
This embodies the requirement that configurations related by a gauge transformation 
are equivalent. We will interpret this in a different way below, and it turns out
to be of crucial relevance to gravity.

Interpretation of the non-Abelian Gauss law: When $A_a^i=0$ we have 
$\int_S dS_a E^{ai}=0$ for any closed surface $S$. 
$dS_a$ is an area element in the local coordinates used.
Associate a $\it physical$ area $dA^i=dS_a E^{ai}$ to this area bit.The
divergence-free condition means that the vectorial sum of the physical areas
is zero, i.e., the surface bits fit together to form a closed surface when 
described in terms of their physical areas also.

To clarify the interpretation when $A_a^i \neq 0$, consider a lattice in
$\Sigma$ formed by an arbitrary grid of the coordinate axes $x^a, a=1,2,3$ close 
together.
Consider a face spanned by the vertices 
\begin{equation}\label{face}
\vec{x}, \vec{x} +\Delta x^1 \vec{n}_1, \vec{x}+ \Delta x^2 \vec{n}_2,
\vec{x} + \Delta x^1 \vec{n}_1+ \Delta x^2 \vec{n}_2
\end{equation}
where $ \vec{n}_1, \vec{n}_2$ are unit vectors along the positive $1$- and $2$- 
axes and $\Delta x^a > 0$.  Associate a physical area of magnitude
$\Delta x^1 \Delta x^2 |\vec E^3(\vec{x})|$ to this face. However assign 
different orientations to it in the frames used in the two cells on either
side of it. Specifically, assign vectorial areas
\begin{equation}\label{area}
{\cal E}_{\pm}^{3i}=\pm \Delta x^1 \Delta x^2 (exp(\pm \frac{1}{2}
 \Delta x^3 \vec{A}_3 . \vec T))^i_j E^{3j}
\end{equation}
in the cells to its positive and negative side
respectively. Here $(\vec{T}^i)_{jk}=-\epsilon_{ijk}$ are the anti-symmetric 
matrices denoting the generators 
in the adjoint representation of $SU(2)$. An overall minus sign for 
$\vec{ \cal E}_{-}$ in (\ref{area}) is a reflection of the fact that the 
outward normal (with respect to the appropriate
cell ) is used to describe the area element.
Thus the same area element is rotated by the orthogonal matrix $\cal O$,
\begin{equation}\label{link}
 {\cal O}= exp(\Delta x^3 \vec A_3 . \vec T)
\end{equation}
in the two adjoining cells. Now demand that the physical areas of the 
six faces of any cell, when all are described by the frame used for it,
vectorially add up to zero. Thus the faces close in the
physical space also. This requirement gives the non-Abelian Gauss law, 
\ref{gauss} in the limit when the grid is infinitely fine. 
It is natural to associate ${\cal E}_{-}^{3i}$ with the dual coordinate
$\vec{x}^{*}$ pertaining to the cell and ${\cal E}_{+}^{3i}$ with
the dual coordinate $\vec{x}^{*}+ \Delta x^{3} \vec{n}_3$
pertaining to the neighboring cell in the third direction. Similar
definitions for other directions are obvious. 
Now the Gauss law \ref{gauss} can be rewritten as,
\begin{equation}\label{lgauss}
{\cal E}_{+x}(\vec{x}^{*}) + {\cal E}_{-x}(\vec{x}^{*}) + 
{\cal E}_{+y}(\vec{x}^{*}) + {\cal E}_{-y}(\vec{x}^{*}) +
{\cal E}_{+z}(\vec{x}^{*}) + {\cal E}_{-z}(\vec{x}^{*})=0
\end{equation}
This form allows an explicit solution of the Gauss law in the quantized case as will 
be shown later.

Our procedure has an obvious relation to lattice gauge theory \cite{lgt}.
The rotation matrix (\ref{link}) is associated with the link joining the vertices
$ \vec{x}^*$ and $ \vec{x}^{*} + \Delta x^{3} \vec{n}_3$ of the dual lattice and
is the holonomy element associated with this infinitesimal line element. 
The electric field $\vec E^{3}(\vec{x})$ is formally associated with the midpoint
of the associated dual link  $(\vec{x}^{*},\vec{x}^{*}+ \Delta x^{3} \vec{n_3})$.
The new variables $\vec{ \cal E}_{\pm}$ are the 
electric fluxes on the face (\ref{face}) of our grid, `parallel transported' to the two
vertices of the dual link, $\it via$ the connection $\vec A$,
to the negative and positive side respectively.
The factor $1/2$ in the exponent in (\ref{area}) is because the parallel transport is by
only half the length of the dual link. Thus all the variables in the Gauss law
(\ref{lgauss}) transform under gauge transformations as vectors at the dual 
coordinate $\vec x^{*}$ labeling the cell under consideration.
This procedure of splitting the link into two makes the definition of the
${\cal \vec{E}}_{\pm}$ variables (\ref{area}) symmetric, but would appear to be 
inessential to interpret the Gauss law. We
could have adopted a convention of associating $\vec{E}$ to one of the 
associated dual vertices, and defined the other by a parallel transport through
the entire dual link to the other dual vertex. However we will see 
that our definition is significant when we consider 
the Poisson bracket algebra of the variables \ref{area}. It turns out that
they can be interpreted as the body- and the face-fixed angular momentum
variables of lattice gauge theory \cite{lgt}.

The interpretation of a lattice regularization is not essential.
As emphasized in \cite{rov} we may interpret it as a floating lattice in the continuum 
theory, used to obtain the later by a limiting procedure.
The coordinate grid may be viewed as a special graph convenient to take
the limit.

It is to be noted that if  ${\cal \vec{E}}_{\pm}$ are specified for any surface bit,
the corresponding $\vec{A}$ on the associated dual link 
is not completely determined but it will have only two degrees of freedom. This 
is as it should be, because the Gauss law (\ref{gauss}) will not constrain
the configuration space completely. It is also important to note that  though the 
cell has the shape of a cuboid in the local coordinates used, it cannot
be assigned a definite  shape in the physical space.In the frame relevant to one cell, 
it can be given such a shape by computing
the vectors corresponding to the edges from the   ${\cal \vec{E}}_{\pm}$. However
these lengths will not match in the neighbouring cubes.(This is discussed further
below. ) This discussion is to emphasize that
the area bits emerge as the crucial entities, and not other objects.
To make the length bits the relevant objects we would need the non-Abelian Stokes law
as is discussed below.

Interpretation of the non-Abelian Stokes law: the non-Abelian analogue of the
Stokes law is,
\begin{equation}\label{stokes}
\epsilon_{abc} D_{b}[\vec{A}] \vec{e}_c(\vec{x})=0.
\end{equation}
This is encountered as the condition that the dreibein
$ e_a^i(\vec{x})$ be torsion-free with respect to the connection 1-form
$ A_a^i(\vec{x})$ . This too has a geometric meaning.
When $A_a^i=0$ we have $\int_C dx^a e_a^i(\vec{x})=0$ for any closed loop $C$. 
$dx^a$ is the line element of the loop in the local coordinates used.
Associate a $\it physical$ length vector $ dl^i=dx^a e_a^i$ to this length bit.The
curl free condition means that the vectorial sum of the physical lengths
is zero, i.e., the length bits fit together to form a closed loop when 
described in physical lengths also.

When $A_a^i \neq 0$, again associate a $\it physical$ length of magnitude
$| \Delta l_a|=\Delta x^a |\vec e_a|$  (where there is no sum over $a$)
with each side $ \Delta x^a$ of the lattice. First we associate a different frame 
for each $\it face$ of the lattice.
The above line element is assigned different orientations
in each of the frames used for the six faces adjoining it.
For the face (\ref{face}), the edges $(\vec{x},\vec{x} + \Delta x^a \vec{n}_a)$, 
(where $a=1,2$ and there is no sum over $a$)
attached to the lowest coordinate $\vec{x}$ of this face
are assigned directly the physical length vectors
$\Delta_a \vec{l} =\Delta x^a \vec{e}_a$.
However for the other two edges a parallel transport is required to compare
with these vectors in the frame relevant to this face. For ex, the edge
$(\vec{x} + \Delta x^1 \vec{n}_1,
 \vec{x}+\Delta x^1 \vec{n}_1+ \Delta x^2 \vec{n}_2)$
is associated with the length vector
$\Delta_2 l^i =\Delta x^2 (exp(\Delta x^1 \vec A_1 . \vec T))^i_j e_2^{j}$
obtained by a parallel transport along the 1-axis to the point $\vec x$.

In this case it is also possible to choose orientations consistently for each
length bit in a frame chosen for each $\it cell$, so that the length bits (
and hence the areas computed from them in the frame) close. The asignment is
as follows: for each of the length bits in the three faces attached to the
`lowest' coordinate of the cell, the orientations are as chosen above. For the
length bits on the remaining three faces of the cell, `parallel transport' all
the orientations chosen on the face using the connection on the link connecting
the face to the `lowest' coordinate.

This gives the relationship to the Regge calculus. In this case the 
physical lengths assigned give a consistent simplicial complex.
Specification of $e$'s in the frame of each
face now completely specifies the $A$'s. This corresponds to the fact that the
torsion-free condition can be used to solve for $A$ algebraically as a local
expression in $e$'s. In this case, the faces are built out of specified length
bits in the physical space also.

All this is in contrast to the case of the Gauss law. Consider the
problem of constructing the areas out of length bits. Given the
configuation of $\vec E$s satisfying the Gauss constraint (\ref{gauss})
with a specified connection $\vec A$,
we may construct a dreibein $\vec e$ by  ${\vec E}^{a}=\epsilon_{abc}
{\vec e}_{b} \times {\vec e}_{c}$. We may also construct the corresponding
connection ${\vec A}[{\vec E}]$ which is torsion-free with respect to which this 
dreibein. Generically,  this will not coincide with the specified connection $\vec A$.
In this case we get frames  in each cell,
such that the areas of the faces computed in the frame coincide precisely with
the configuation of $\vec E$s. But these frames will be related to 
${\vec A}[{\vec E}]$  and not to the specified connection $\vec A$.
Therefore the area vectors computed will not have the required orientations,
though they have the correct magnitude.

The area variables we have defined have a very significant Poisson brackets 
algebra among each other. With our discretiztion,
$\delta(\vec{x}) \rightarrow (\Delta x^1 \Delta x^2 \Delta x^3)^{-1}$.
Therefore, $[\Delta x^3 A_3^i(\vec{x}),\Delta x^1 \Delta x^2 E^{3j}(\vec{y})]_{PB}=
\delta_{ij} \delta_{\vec{x}, \vec{y}}$. Thus the variables in the bracket form  
canonically conjugate pairs.

Poisson bracket algebra: We now compute the Poisson bracket,
\begin{eqnarray*}
[ {\cal E}_{+}^{3a} (\vec{x}^{*}), {\cal E}_{\pm}^{3b} (\vec{y}^{*})]_{PB}=
\pm \Delta x^1 \Delta x^2 E^{3k} 
(- \epsilon_{lik} (exp(\pm \frac{1}{2} \Delta x^3 \vec A_3 . \vec T))^j_l 
\pm \epsilon_{ljk} (exp(\frac{1}{2} \Delta x^3 \vec A_3 . \vec T))^i_l) 
\end{eqnarray*}
Expanding the exponent in powers of $\Delta x^{3}$ we see that
$({\cal E}_{+}^{3a}, a=1,2,3)$ has the algebra of angular momentum among each
other for a given cell whereas they  commute with $({\cal E}_{-}^{3b}, b=1,2,3)$ 
even in the same cell, to order $O(\Delta x^3)$.
Thus in the limit of an infinitely fine grid,
${\vec {\cal E}}_{\pm}^{ai} (\vec{x}^{*})$ for each 
$\pm$, $\vec{x}^{*}$ and $\vec{a}$ have the PB of angular momentum, whereas they have
zero PBs with all other variables.This identifies them with the body- and space 
fixed angular momenta of $SO(3)$ lattice gauge theory, which generate an $SO(3)$
rotation of the link variable on the left and right respectively \cite{lgt}. In fact 
$\vec{\cal E}_{\pm} (\vec{x}^{*})$ respectively
generate an $SO(3)$ rotation of all Wilson lines starting from the
point and having the tangent along the positive and negative 3-direction 
respectively at this point. Our procedure of splitting the dual link
makes this operation precise. If the Wilson line begins with the positive
(negative) half of the dual link, it transforms non-trivially under
$\vec{\cal E}_{\pm} (\vec{x})$ respectively. In effect $ \vec{E} (\vec{x})$ in
${\vec{\cal E}}_{\pm} (\vec{x}^{*})$ (see (\ref{area})) removes the orthogonal matrix 
associated with this half link and it is replaced back
by the same orthogonal matrix contained in $\vec{\cal E} _{\pm}$. 
 
Quantization: With this Poisson bracket structure, prescription for quantization 
is evident. Each ${\cal E}_{+}^i$ or ${\cal E}_{-}^i$ becomes the three
components of independent angular momentum operators. There is however the
constraint that  
$\vec{\cal E}_{+}^{2} (\vec{x}^{*}+ \Delta x^3 \vec{n}_3)= 
\vec{\cal E}_{-}^{2} (\vec{x}^{*})$ as the two vectors
differ from each other only by an orthogonal rotation. 
This condition is easy to impose. Ignoring this condition
we have the standard basis 
$|j_{+}m_{+}>|j_{-}m_{-}>$ where the total angular momentum and the `third'
component are simultaneously diagonalised.Now the constraint is simply
solved by choosing a subset of this basis with $j_{+}=j_{-}(=j$ say).
Thus the label for the basis is $|j,m_{+},m_{-}>$. The reason for this is
simple. The link variable may be interpreted \cite{lgt} as the configuration of a 
rigid body described by the  rotation $\cal O$ from
the space- to the body- fixed axes.  The corresponding wave function
$\Psi(\cal O)$ can be expanded in the basis $D^j_{mn}$ of the rotation matrices.
$j$ labels the total angular momentum which is the same in the two frames.
$m,n$ label the eigenvalues of the `third component' of the body- and space-
fixed angular momenta which can be simultaneously diagonalised.

Gauss law in form (\ref{lgauss}) is simply the constraint that the six angular momenta
associated with the six faces of a cell add up to zero. In quantum theory 
this can be solved by going to an appropriate coupled basis.
This enables us to construct a complete orthonormal basis which satisfies 
the Gauss law. This procedure is a simple extension of the one used in \cite{2lgt} for
lattice gauge theory in two space dimensions, where the basis was interpreted
in terms of triangulations. The main difference is that the ${\cal E}$s are now 
interpreted as area bits instead of the length bits.

We choose a convention for the sequence in which the angular momenta are added. Any
other convention leads to an equivalent basis, though some of them may not lead to the
geometric interpretation given below.For each 3-cell, we add the angular momentum
pair-wise in the following sequence repeatedly:

\begin{eqnarray*}
|j_{+x} m_{+x}> |j_{+y}m_{+y}> |j_{+z}m_{+z}>
|j_{-x} m_{-x}> |j_{-y}m_{-y}> |j_{-z}m_{-z}> \\
\longrightarrow
|j_{+x}j_{-y}j_{+x-y}m_{+x-y}>
|j_{+y}j_{-z}j_{+y-z}m_{+y-z}>
|j_{+z}j_{-x}j_{+z-x}m_{+z-x}> \\
\longrightarrow
|j_{+x}j_{-y}j_{+y} j_{-z} j_{+x-y} j_{+y-z} j_{+x-y+y-z} m_{+x-y+y-z}>
|j_{+z}j_{-x}j_{+z-x}m_{+z-x}> \\
\longrightarrow
|j_{+x}j_{-y}j_{+y}j_{-z}j_{+z}j_{-x}j_{+x-y}j_{+y-z}j_{+z-x} j_{+x-y+y-z} 
j_{+x-y+y-z+z-x} m_{+x-y+y-z+z-x}>
\end{eqnarray*}

In the basis written last it is easy to implement the Gauss constraint. It simply
gives
$j_{+x-y+y-z+z-x}=m_{+x-y+z-x+y-z}=0$. In this case we have necessarily,
$j_{+x-y+y-z}=j_{+z-x}$.
Thus the complete orthonormal basis satisfying the Gauss law (\ref{gauss})
is simply labeled by the integer labels,
$|\{j_{+x}(\vec{x}^{*}) j_{+y}(\vec{x}^{*}) j_{+z}(\vec{x}^{*}) 
j_{-x}(\vec{x}^{*}) j_{-y}(\vec{x}^{*}) j_{-z}(\vec{x}^{*}) 
j_{+x-y}(\vec{x}^{*}) j_{+y-z}(\vec{x}^{*}) j_{+z-x}(\vec{x}^{*}) \}>$.
It is further understood that
$j_{-a}(\vec{x}^{*})= j_{+a}(\vec{x}^{*} + \Delta x^{a} \vec{n}_{a})$
for each $a=1,2,3$.

The interpretation is follows: $j$s are the magnitudes of the areas associated with the
faces of our cells.They are quantized. Further there are constraints.
For example, $(j_{+x}, j_{-y}, j_{+x-y})$ pertaining to one cell satisfy the
triangle rule for the addition of angular momenta. This is interpreted to mean that with
suitable choice of orientations, the area elements associated with $j_{+x}$
and $j_{-y}$ add up vectorially to the area element associated with $j_{+x-y}$.
Again $(j_{+x-y},j_{+y-z},j_{+z-x})$  satisfy the triangle rule.
This is again interpreted to mean that corresponding areas close on appropriate
choice of their orientations. The orientations assigned for (say) $j_{+x-y}$  
in these two cases need not match.

It is also possible to interpret the Gauss constraint in the present case
in terms of length bits \cite{3lgt}. These form the sides of truncated
octohedrons which are stacked end-to end.

Diffeomorphism invariance: We now consider the implications of the next three
constraints \cite{bar}. They imply that the theory is invariant under 
diffeomorphisms of 3-dimensional space. Gauss law (\ref{gauss}) is form invariant
under an arbitrary change of the local coordinates $\vec x$ provided $\vec A$
and $\vec E$ are appropriately transformed. Consider the coordinate grid with the
new coordinates. The new $\vec{\cal E}$s are related to the earlier ones 
(in \ref{area}) as follows: consider the electric flux on the faces of the new grid
instead of the old one. These are related to the earlier ones by a linear 
transformation. Also, the orthogonal matrix (\ref{link})  to be used 
now pertain to the 
holonomy elements along the new dual links. The appropriate Lie algebra
element (i.e. the exponent in (\ref{area})) is again a linear combination of the previous 
ones.The three space diffeomorphism constraints imply that these two configurations
are physically equivalent.  The way to handle this and its implications will be
considered elsewhere.

We now briefly comment on the continuum limit. This corresponds to the
limit of an infinite number of cells. It is to be expected that the discreteness
of area does not get washed out in this limit at least for certain macroscopic
situations such as the black holes.

Our considerations suggest that quantum gravity is a theory of area bits with 
quantized areas. The magnitudes of these areas can be simultaneously
observed, but the orientations are indefinite, due to the quantum mechanical 
uncertainty. However, the magnitudes are such that it is possible to fit them 
together into closed surfaces with some choice of orientations. 
This brings out the specific way in which classical geometry is relevant to 
quantum gravity. As was noted before, though these area bits were associated with a
rectangular bit
in the coordinates chosen to describe the manifold, they cannot be associated with
specified edges in physical space. As discussed, such a specification would be a
consequence of the torsion-free condition instead of the Gauss constraint.
This may be again interpreted as a consequence of the quantum mechanical uncertainty which disallows a
simultaneous specification of the conjugate variables.

The geometry relevant to the basis of states obtained here will be addressed in more 
detail elsewhere \cite{long}. 

We thank Professor Ramesh Anishetty for discussions on issues and techniques
in gravity and gauge theories.

\end{document}